\documentclass[aps,showpacs]{revtex4}
\usepackage{graphicx}
\usepackage[all]{xy}
\usepackage{amsmath}
\usepackage{amssymb}
\usepackage[colorlinks=true, pdfstartview=FitV, linkcolor=blue, citecolor=red, urlcolor=magenta]{hyperref}
\usepackage[usenames,dvipsnames]{color}
\usepackage[utf8]{inputenc}

\newcommand{\bes}{\begin{subequations}}
\newcommand{\ees}{\end{subequations}}
\def\ben{\begin{eqnarray}}
\def\een{\end{eqnarray}}
\def\be{\begin{equation}}
\def\ee{\end{equation}}

\begin{document}
\title{New family of potentials with analytical twiston-like solutions}
\author{J. R. L. Santos$^{1}${\footnote{Electronic address: joaorafael@df.ufcg.edu.br }}, D. S. S. Borges$^{1}${\footnote{Electronic address: damaresborges@hotmail.com }}, and I. O. Moreira$^{1}${\footnote{Electronic address: ivandersonoliveira@hotmail.com }}}
\affiliation{$^1${Unidade Acad\^emica de F\'\i sica, Universidade Federal de Campina
Grande, Caixa Postal 10071, 58109-970, Campina Grande PB, Brazil} }
%\affiliation{$^2${Departamento de F\'\i sica, Universidade Federal da Para\'iba, Caixa Postal 5008, 58051-970 Jo\~ao Pessoa PB, Brazil} }
%\date{\today}
\begin{abstract}
In this letter we present a new approach to find analytical twiston models. The effective two-field model was constructed by a non-trivial combination of two one field systems. In such an approach we successfully build analytical models which are satisfied by a combination of two defect-like solutions, where one is responsible to twist the molecular chain by $180^{\,0}$, while the other implies in a longitudinal movement. Such a longitudinal movement can be fitted to have the size of the distance between adjacent molecular groups. The procedure works nicely and can be used to describe the dynamics of several other molecular chains.
\end{abstract}

\maketitle

%%%%%%%%%%%%%%%%%%%%%%%%%%%%%%%%%%%%%%%%%%%%%%%%%

\section{Introduction}
\label{intro}

In the end of seventies Mansfield and Boyd \cite{mansfield78} introduced the concept of twist-like solutions or twistons, as a $180^0$ twist of several $CH_2$ groups in a crystalline polyethylene molecule. The twiston emerges as consequence of the large torsional flexibility of the crystalline chain, and it appears in the plane orthogonal to the chain direction. Since the seminal work of Mansfield and Boyd, solutions like twistons are part of different branches in the scientific literature, covering subjects as organic chains \cite{su79}, or even as the description of deformations in nanoribbons \cite{savin17} and also as consequence of anharmonic interactions in nanotubes \cite{savin}.

Most part of the applications of twistons are based on well-known soliton solutions or in numerical calculations. In the search for a different analytical approach of the twistons proposed by \cite{mansfield78}, Bazeia and collaborators \cite{bazeia99,bazeia00} introduced a methodology based on classical field theory, which remarkably describes the expected behavior of a twist-like model. The model described in \cite{bazeia99,bazeia00} also presents an infinite number of degenerated states, due to the form of its potential.

An attempt to break such a degeneracy was proposed by \cite{dutra10}, where the authors added a perturbation function to the interaction potential. Despite the fact that such a perturbation function broke the degenerated states, it yielded to equations of motion which were non-longer analytical. Therefore, only numerical twistons were found in this last approach. In this study we introduce a new route to treat the degeneracy issue, keeping the twist-like solutions analytical. Such a methodology is grounded in the so-called extension method, and the effective twiston model is build naturally. 

The discussions tailored in this letter are divided as follows: Firstly we review some generalities about the analytical treatment of twistons. After that, we present the general aspects about the extension method. There we explain carefully how the method works and how it can be implemented in the context of twiston-like models. Then, we construct a new twiston model by applying the extension method, and we derive a new family of potentials which share the same analytical solutions. Finally we present our final remarks and future perspectives.

\section{Generalities}
\label{gen}
We begin this generalities by reviewing the connection between molecular dynamics and classical field theory previously investigated by \cite{bazeia99,bazeia00,dutra10}. Let us firstly assume that we are dealing with a rigid molecular group, which can be described through cilindrical coordinates. Therefore, the kinetic energy of the chain is such that
\be
T=\frac{1}{2}\,m\,r_0^{\,2}\,\sum_{n}\,\left[\dot{\phi}_n^{\,2}+\frac{c^{\,2}}{r_0^{\,2}}\,\dot{\chi}_n^{\,2}\right]\,,
\ee
here $r_0$ is the equilibrium position of the radial coordinate, $c$ is the longitudinal distance between adjacent molecular groups, and we are neglecting the radial degree of freedom. 

Thus, a Lagrangian formalism for such a chain can be written as
\be
L=T-U(\phi_{n},\chi_{n})\,,
\ee
where $U$ is the intermolecular potential. In the continuum limit the coordinates $\phi_n$ and $\chi_n$ are going to behave like real fields $\phi(x,t)$ and $\chi(x,t)$. Therefore, the  twistons' dynamics can be modeled through a classical relativistic Lagrangian composed by these two scalar fields, whose explicit form is
\be
\label{gen_01}
{\cal L}=\frac{1}{2}\,\partial_{\,\mu}\phi\,\partial^{\,\mu}\phi+\frac{1}{2}\,\partial_{\,\mu}\chi\,\partial^{\,\mu}\chi-V(\phi,\chi)\,,
\ee
as discussed by \cite{mansfield78, zhang94, bazeia99, bazeia00, dutra10}. Here $\mu=0,1$ but we are going to work with statical fields, which do not change the physical properties of the model, since time dependent fields can be derived taking a Lorentz boost. In order to find analytical topological defects, as the agents of the twist movement, we assume that the potential is positive defined and written in terms of a superpotential such as
\be
\label{gen_02}
V(\phi,\chi)=\frac{1}{2}\,W_{\,\phi}^{\,2}+\frac{1}{2}\,W_{\,\chi}^{\,2}\,,
\ee
where, $W_{\,\phi}=\partial\,W/\partial\,\phi$, and $W_{\,\chi}=\partial\,W/\partial\,\chi$. If we deal with statical one-dimensional fields, the total energy related with these defects is given by
\ben
\label{gen_03} 
E &=&-\int_{-\infty}^{+\infty}\,dx\,{\cal L} \\ \nonumber
&=&
\int_{-\infty}^{+\infty}\,dx\,\left[\frac{1}{2}\,\left(\frac{d\,\phi}{d\,x}\right)^{\,2}+\frac{1}{2}\,\left(\frac{d\,\chi}{d\,x}\right)^{\,2}+V(\phi,\chi)\right]\,.
\een
Let us make use of the so-called BPS method (Bogomol'niy-Prasad-Sommerfield) \cite{bps,bps2}, to rewrite the total amount of energy as
\be
\label{gen_04}
E=\frac{1}{2}\,\int_{-\infty}^{+\infty}\,dx\,\bigg[\left(\frac{d\,\phi}{d\,x}\mp W_{\,\phi}\right)+ \left(\frac{d\,\chi}{d\,x}\mp W_{\,\chi}\right)
\pm 2\,\left(\frac{d\,\phi}{d\,x}\,W_{\,\phi}+\frac{d\,\chi}{d\,x}\,W_{\,\chi}\right)\bigg]\,.
\ee
Therefore, imposing the first-order differential equations
\be
\label{gen_05}
\frac{d\,\phi}{d\,x}=W_{\,\phi}(\phi,\chi)\,; \qquad \frac{d\,\chi}{d\,x}=W_{\,\chi}(\phi,\chi)\,,
\ee
as constraints which minimize the energy, we yield to the well-known BPS energy
\be
\label{gen_06}
E_{BPS}=\left|\Delta\,W\right|\,; \qquad \Delta\,W= W_{+\infty}-W_{-\infty}\,,
\ee
where
\be
W_{+\infty}=W\left[\phi(x(+\infty)),\chi(x(+\infty))\right]\,; \qquad
W_{-\infty}=W\left[\phi(x(-\infty)),\chi(x(-\infty))\right]\,.
\ee
Here we point that the solutions from the coupled differential equations presented in $(\ref{gen_05})$ also satisfy the equations of motion coming from $(\ref{gen_01})$. 

\begin{figure}
 \includegraphics[width=0.45\columnwidth]{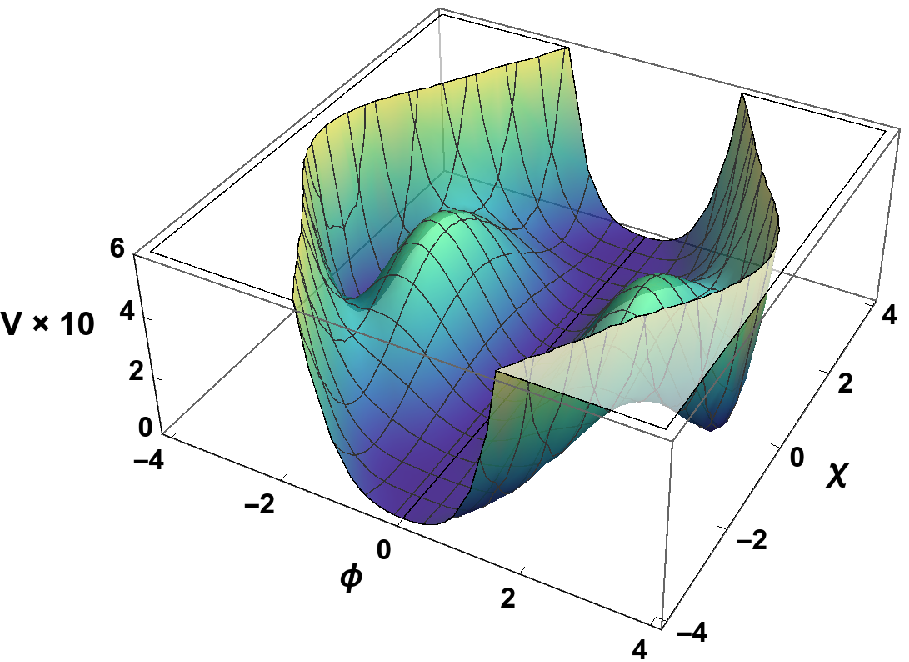} \hspace{0.5 cm} \includegraphics[width=0.45\columnwidth]{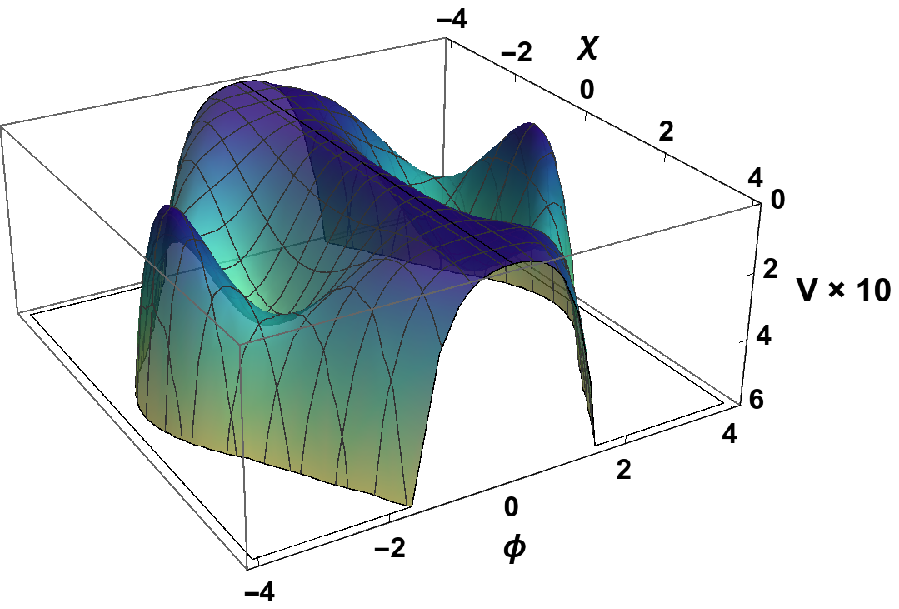}
 \caption{Potential $V$ for $\lambda=7.30\,\pi^{\,-2}$, $\mu=0.67$. The graphics show a line of zeros for $\phi=0$, besides two other vacua in $\phi=\pm\pi$ and $\chi=0$.}
 \label{FIG1}
\end{figure}

One possible model introduced by Bazeia and collaborators \cite{bazeia99,bazeia00}, and also studied in \cite{dutra10} consists in the potential
\be
\label{gen_07}
V=\frac{1}{2}\,\left[\lambda\,\phi\,(\phi^{\,2}-\pi^{\,2})+\mu\,\phi\,\chi^{\,2}\right]^{\,2}+\frac{1}{2}\,\left(\mu\,\phi^{\,2}\,\chi\right)^{\,2}\,,
\ee
where $\lambda$ and $\mu$ are real constants. An interesting point to be noted about this model is that the potential $V$ presents a line of zeros when $\phi=0$, as one can see in Fig. \ref{FIG1}. The first-order differential equations for such a potential are
\be
\label{gen_08}
\frac{d\,\phi}{d\,x}=\lambda\,\phi\,(\phi^{\,2}-\pi^{\,2})+\mu\,\phi\,\chi^{\,2}\,; \qquad \frac{d\,\chi}{d\,x}=\mu\,\phi^{\,2}\,\chi\,,
\ee
yielding to the analytical twistons
\be
\label{gen_09}
\phi(x)=\pm\,\pi\,\sqrt{\frac{1}{2}\,\left(1-\tanh(\mu\,\pi^{\,2}\,x)\right)}\,; \qquad \chi(x)=\pm\,\pi\,\sqrt{\frac{\lambda}{\mu}-1}\,\sqrt{\frac{1}{2}\,\left(1+\tanh(\mu\,\pi^{\,2}\,x)\right)}\,,
\ee
which were calculated in \cite{bazeia99,bazeia00}. In this last expression we realize that field $\phi$ is responsible for the torsion motion of the polymer chain, while field $\chi$ is equivalent to its longitudinal motion. We also see that the previous solutions are valid only if $\lambda/\mu\,>\,1$, as pointed by \cite{bazeia99,bazeia00}. 

The superpotential used to described this analytical model is
\be
\label{gen_10}
W(\phi,\chi)=\frac{\lambda}{2}\,\phi^{\,2}\,\left(\frac{\phi^{\,2}}{2}-\pi^{\,2}\right)+\frac{\mu}{2}\,\phi^{\,2}\,\chi^{\,2}\,,
\ee
which is constructed from the first-order differential equations $(\ref{gen_05})$. Then, we substitute the superpotential together with the analytical solutions $(\ref{gen_09})$ into Eq. $(\ref{gen_06})$, yielding to the BPS energy
\be
\label{gen_11}
E_{BPS}=\frac{\lambda}{4}\,\pi^{\,4}\,.
\ee
This result was derived in \cite{bazeia00}, where the authors worked with $|\lambda|\,\pi^{\,2}=7.30\,\mbox{kcal/mol}$. Such a value was based on the experimental data for the energy contribution of the localized twisted region \cite{mansfield78}. This experimental constraint yields to $E=17.99\,\mbox{kcal/mol}$ as the energy for the topological twistons. As mentioned in \cite{bazeia00}, this energy agrees quite well with other methods to describe twistons as one can see in \cite{mansfield78, skinner84, zhang94}.

There is one important point to be noted about the line of zeros presented in Fig. \ref{FIG1}, which was discussed in details by de Souza Dutra {\it et al.} in \cite{dutra10}. As was shown before, field $\phi$ is responsible to twist the molecule chain by an integer multiple of $\pi$, while field $\chi$ produces a longitudinal unit shift for the chain, see $(\ref{gen_09})$. Such a longitudinal shift should be of order of the longitudinal distance between adjacent molecular groups. However, once the line of zeros presents degenerated energy for any value of $\chi_v$, this means that it would be possible to break the chain without any extra energy cost than that necessary to make an unit shift \cite{dutra10}.

In reference \cite{dutra10}, the authors proposed a route to break this degenerated energy configuration, by adding a perturbation function to the original potential. The procedure worked nicely, however the solutions for the fields should be numerically derived.  In order to keep an analytical description for the twistons, we are going to apply the so-called extension method \cite{bls} to build another two-field models which are going to have well defined vacua to field $\chi$. This methodology rescues the physical meaning of the analytical twistons derived in \cite{bazeia00,bazeia99}, despites the degeneracy issue.    

\section{The extension method}
\label{extmet}

In this section we present some generalities about the extension method firstly introduced in \cite{bls}. In order to apply such a method, let us verify that the first-order differential equations from \eqref{gen_05} can be rearranged as
\be \label{sec2_eq1}
\phi_{\,\chi}=\frac{\phi^{\,\prime}}{\chi^{\,\prime}}=\frac{W_{\,\phi}(\phi,\chi)}{W_{\,\chi}(\phi,\chi)}\,,
\ee
whose integration lead us to analytical orbits between $\phi$ and $\chi$, or in other words, we find $\phi=f(\chi)$. Such a mapping procedure is approached in a different context by the so-called deformation method \cite{blm}. 

In the deformation method, we work with two standard classical field Lagrangians, whose forms are
\be \label{sec2_eq2}
{\cal L}=\frac{1}{2}\,\partial_{\,\mu}\,\phi\,\partial^{\,\mu}\,\phi-V(\phi)\,; \qquad {\cal L}_d =\frac{1}{2}\,\partial_{\,\mu}\,\chi\,\partial^{\,\mu}\,\chi-\widetilde{V}(\chi)\,,
\ee
where the real fields $\phi$ and $\chi$ are statical, one dimensional, and are solutions of  the equations of motion
\be \label{sec2_eq3}
-\phi^{\,\prime\,\prime}+V_{\,\phi}=0\,; \qquad -\chi^{\,\prime\,\prime}+\widetilde{V}_{\,\chi}=0\,.
\ee
Moreover, the BPS defects should also satisfy the first-order differential equations
\be \label{sec2_eq4}
\phi^{\,\prime}=W_{\,\phi}(\phi)\,; \qquad \chi^{\,\prime}=W_{\,\chi}(\chi)\,,
\ee
where  $V$ and $\widetilde{V}$ are given by
\be \label{sec2_eq5}
V=\frac{W_{\,\phi}^{\,2}}{2}\,; \qquad \widetilde{V}=\frac{W_{\,\chi}^{\,2}}{2}\,.
\ee
The connection between these two models is mediated by a function $f$, which is named deformation function. Therefore, a connection of type $\phi=f(\chi)$ yields to the constraints
\be \label{sec2_eq6}
W_{\phi}(\phi)=W_{\,\chi}(\chi)\,f_{\,\chi}\,; \qquad \phi_{\,\chi}=f_{\,\chi}=\frac{W_{\,\phi}}{W_{\,\chi}}\bigg|_{\phi\rightarrow\chi}\,.
\ee

The reader can see that the previous relation is very close to the one presented in Eq.  \eqref{sec2_eq1}. The basis for the extension method is that we can rearrange Eq. \eqref{sec2_eq6} to write it as an effective differential equation of a two-field model. The procedure works as follows: let us rewrite  \eqref{sec2_eq6} as
\be \label{sec2_eq7}
\phi_{\,\chi}=\frac{W_{\,\phi}}{W_{\,\chi}}=\frac{a_1\,W_{\,\phi}(\chi)+a_2\,W_{\,\phi}(\phi,\chi)+a_3\,W_{\,\phi}(\phi)+c_1\,g(\chi)+c_2\,g(\phi,\chi)+c_3\,g(\phi)}{b_1\,W_{\,\chi}(\chi)+b_2\,W_{\,\chi}(\phi,\chi)+b_3\,W_{\,\chi}(\phi)}\,,
\ee
where $W_{\phi}(\chi)$, $W_{\phi}(\phi,\chi)$, and $W_{\phi}(\phi)$ are equivalent since they are constructed with the deformation function $f(\chi)$ and with its inverse $\chi=f^{\,-1}(\phi)$. The analogous procedure is repeated for the different forms of $W_{\chi}$, as well as for the extra function $g$. 

This last function is used to connect fields $\phi$ and $\chi$ in the effective two scalar field model. Moreover, the complete equivalence of the previous equation with  \eqref{sec2_eq6} imposes the constraints $a_1+a_2+a_3=1$, $b_1+b_2+b_3=1$, and $c_1+c_2+c_3=0$.

Consequently, once Eq. \eqref{sec2_eq7} has the same structure of Eq. \eqref{sec2_eq1}, we can identify the following terms
\be \label{sec2_eq8}
W_{\,\phi}=a_1\,W_{\,\phi}(\chi)+a_2\,W_{\,\phi}(\phi,\chi)+a_3\,W_{\,\phi}(\phi)+c_1\,g(\chi)+c_2\,g(\phi,\chi)+c_3\,g(\phi)\,; 
\ee
\be \label{sec2_eq9}
W_{\,\chi}=b_1\,W_{\,\chi}(\chi)+b_2\,W_{\,\chi}(\phi,\chi)+b_3\,W_{\,\chi}(\phi)\,,
\ee
where $W_{\,\phi\,\chi}$ and $W_{\,\chi\,\phi}$ must commutate, which means that
\be  \label{sec2_eq10}
W_{\,\chi\,\phi}=W_{\,\phi\,\chi}\,,
\ee
yielding to a constraint for the function $g$, given by
\be  \label{sec2_eq11}
b_2\,W_{\,\chi\,\phi}(\phi,\chi)+b_3\,W_{\,\chi\,\phi}(\phi)=a_1\,W_{\,\phi\,\chi}(\chi)+a_2\,W_{\,\phi\,\chi}(\phi,\chi)+c_1\,g_{\,\chi}(\chi)+c_2\,g_{\,\chi}(\phi,\chi)\,.
\ee
Then, in order to have an unique form of $g$, we must choose either $c_1$ or $c_2$ equals to zero. Another interesting feature about this extension method is that the analytical solutions of the deformed one field systems are going to automatically satisfy the equations of motion of the effective two-field model. Such a methodology was successfully applied to the construction of new two and three scalar fields cosmological models, as one can see in \cite{ms, smvf}.

\section{Extended Twistons}
\label{extwist}

Let us consider two one-field models which obey the first-order differential equations
\be \label{twist01}
\phi^{\,\prime}=W_{\phi}=\mu\,  \phi\,  \left(\phi ^2-\pi ^2\right)\,; \qquad \chi^{\,\prime}=\mu \, \chi\,  \left(\pi ^2-\frac{\chi ^2}{\gamma^{\,2}}\right)\,; \qquad \gamma=\sqrt{\frac{\lambda}{\mu}-1}\,,
\ee
which are satisfied by the analytical defects
\be \label{twist02}
\phi(x)=\pi \, \sqrt{\frac{1}{2} \left(1-\tanh \left(\pi ^2\, \mu\,  x\right)\right)}\,;\qquad \chi(x)=\pi \, \gamma \sqrt{\frac{1}{2}\, \left(\tanh \left(\pi ^2\, \mu \, x\right)+1\right)}\,.
\ee

From the previous equations, we realize that the mapping between fields $\phi$ and $\chi$ is mediated by 
\be \label{twist03}
\phi=f(\chi)=\sqrt{\pi^{\,2}-\frac{\chi^{\,2}}{\gamma^{\,2}}}\,; \qquad \chi=f^{\,-1}(\phi)= \gamma\,\sqrt{\pi^{\,2}-\phi^{\,2}}\,.
\ee
Then, we are able to use these ingredients to write two different and equivalent forms of $W_{\,\phi}$ and $W_{\,\chi}$, respectively, yielding to
\ben \label{twist04}
&&
W_{\,\phi}(\phi ,\chi)=\lambda\,\phi\,\left(\phi^{\,2}-\pi^{\,2}\right)+\mu\,\chi^{\,2}\,\phi\,; \qquad W_{\,\phi}(\phi)=\mu\,\left(\phi^{\,2}-\pi^{\,2}\right)\,\phi\,; \\ \nonumber
&&
W_{\,\chi}(\chi)=\mu\,\chi\,\left(\pi^{\,2}-\frac{\chi^{\,2}}{\gamma^{\,2}}\right) \,; \qquad W_{\,\chi}(\phi ,\chi)=\mu\,\chi\,\phi^{\,2}\,.
\een
Here  $W_{\phi}(\chi)$ and $W_{\chi}(\phi)$ were neglected once they presented terms with rational powers of $\phi$ and $\chi$. The procedure of neglecting these forms is equivalent to choose $a_1=0$ and $b_3=0$. Therefore, choosing $c_1=0$, the constraint \eqref{sec2_eq11} results in
\be \label{twist05}
c_2\,g_\chi=2\,\mu\,\left(b_2-a_2\right)\,\chi\,\phi\,,
\ee
whose a direct integration lead us to
\be \label{twist06}
c_2\,g(\phi,\chi)=\mu\,\left(b_2-a_2\right)\,\chi^{\,2}\,\phi\,; \qquad c_2\,g(\phi) =\mu\gamma^{\,2}\,\left(b_2-a_2\right)\,\left(\pi^{\,2}-\phi^{\,2}\right)\,\phi\,,
\ee
where the last was obtained using $\chi=f^{\,-1}(\phi)$.

\begin{figure}[ht!]
 \includegraphics[width=0.45\columnwidth]{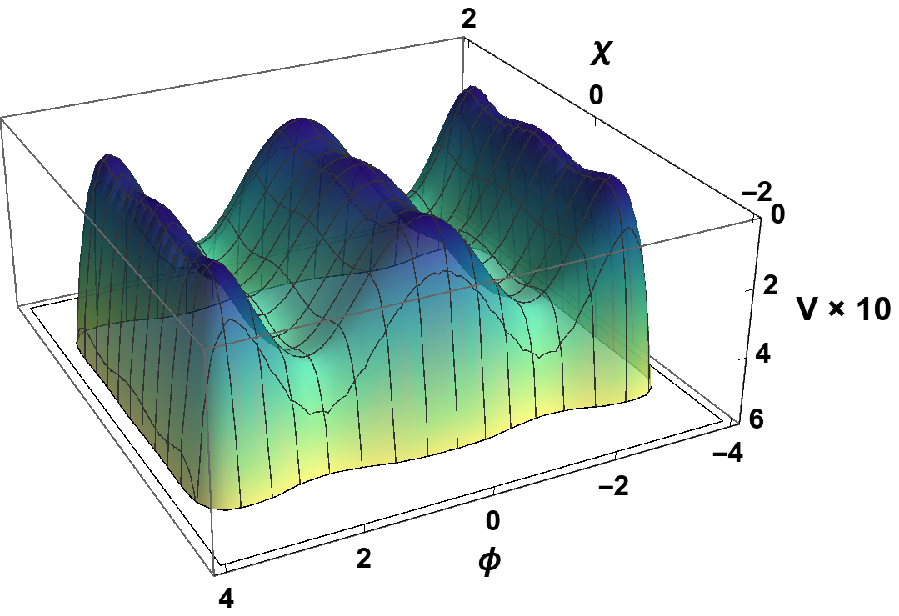} \hspace{0.5 cm} \includegraphics[width=0.45\columnwidth]{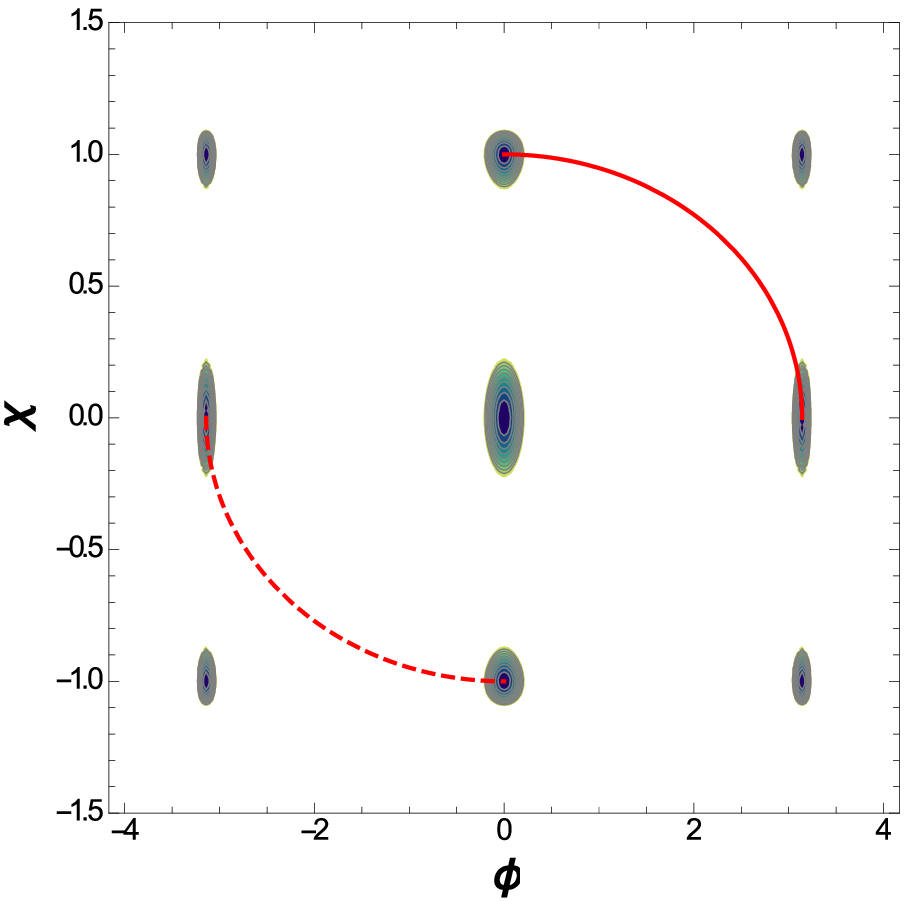}
 \caption{In the left panel we show potential $V$ for $\lambda=7.30\,\pi^{\,-2}$, $\mu=0.67$ and $b_2=0$. In the right panel we present a contour of $V$, were the solid and the dotted red curves are the analytical orbits connecting fields $\phi$ and $\chi$.}
 \label{FIG2}
\end{figure}

\begin{figure}[h!]
 \includegraphics[width=0.45\columnwidth]{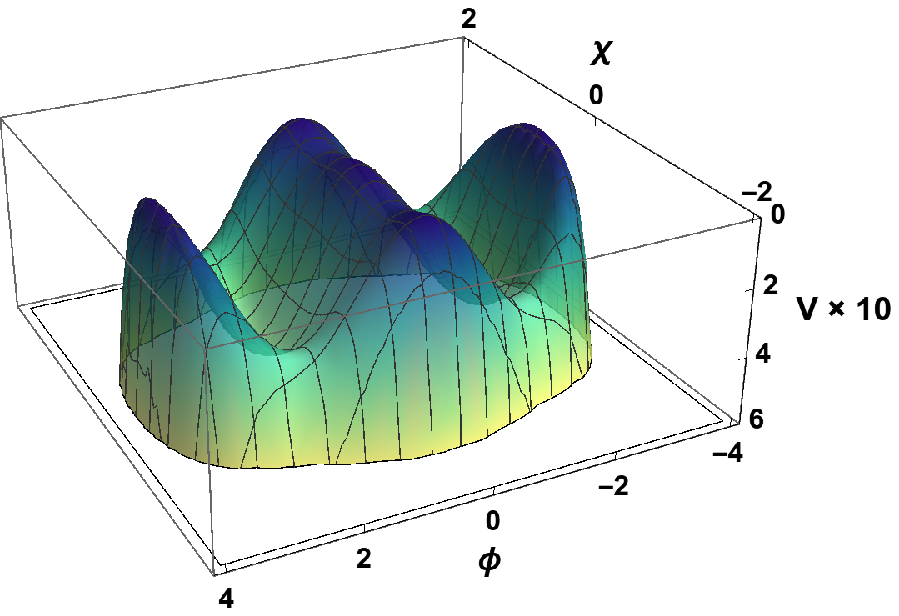} \hspace{0.5 cm} \includegraphics[width=0.45\columnwidth]{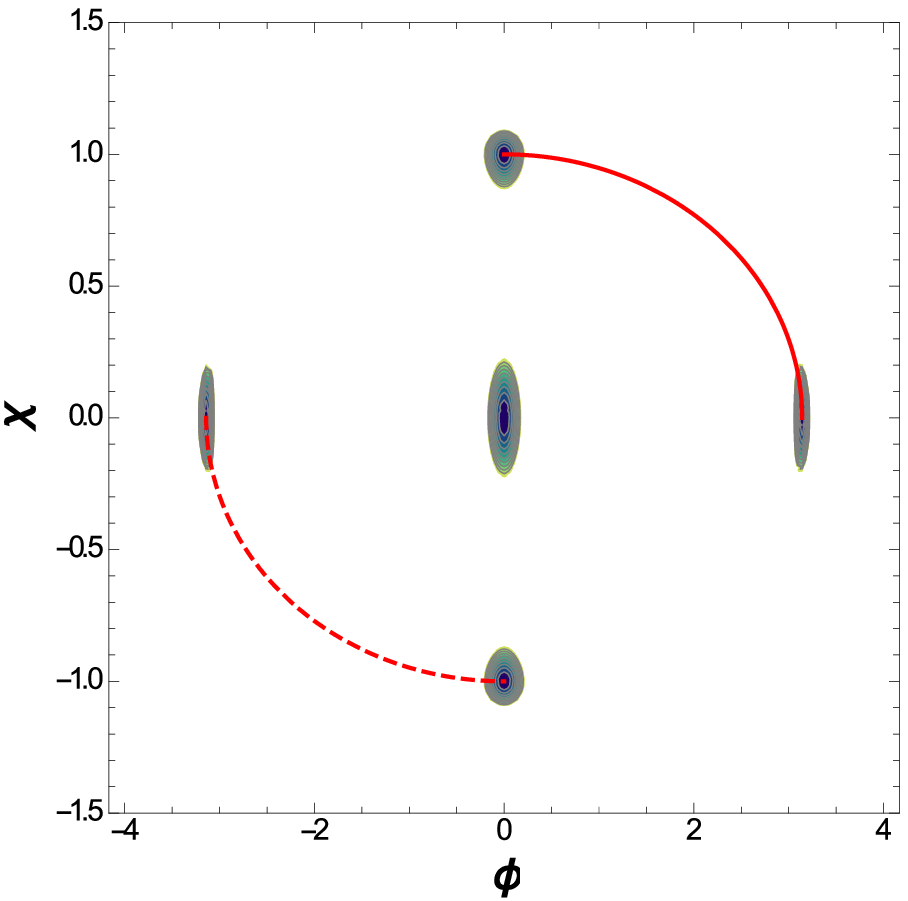}
 \caption{In the left panel we show potential $V$ for $\lambda=7.30\,\pi^{\,-2}$, $\mu=0.67$ and $b_2=2$. In the right panel we present a contour of $V$, were the solid and the dotted red curves are the analytical orbits connecting fields $\phi$ and $\chi$.}
 \label{FIG3}
\end{figure}

By substituting all these ingredients in \eqref{sec2_eq8} and \eqref{sec2_eq9} we find that
\be \label{twist07}
W_{\,\phi}=\left(b_2\,\gamma^{\,2}+1\right)\,\mu\,\left(\phi^{\,2}-\pi^{\,2}\right)\,\phi+b_2\,\mu\,\chi^{\,2}\,\phi\,,
\ee
and
\be \label{twist08}
W_{\,\chi}=(1-b_2)\,\mu\,\left(\pi^{\,2}-\frac{\chi^{\,2}}{\gamma^{\,2}}\right)\,\chi+b_2\,\mu\,\chi\,\phi^{\,2}\,.
\ee
The integration of $W_{\phi}$ and of $W_{\chi}$ in respect to $\phi$ and $\chi$, respectively results in the effective superpotential
\be \label{twist09}
W(\phi,\chi)=\frac{1}{2}\, (1-b_2)\, \mu\, \left(\pi^{\,2}-\frac{\chi ^2}{2\, \gamma^{\,2}}\right)\,\chi ^2+\frac{1}{2} \,\left(b_2\,\gamma^{\,2}+1\right)\,\mu\, \left(\frac{\phi ^2}{2}-\pi ^2\right)\, \phi ^2 +\frac{1}{2}\,b_2\, \mu\,  \chi ^2 \,\phi ^2\,.
\ee
Moreover, from Eq. \eqref{gen_02} we can derive an analytical expression for potential $V$ given by
\ben \label{twist10}
V(\phi,\chi) &=& \frac{\mu^{\,2}}{2}\,\phi^{\,2}\,\left[\left(b_2\,\gamma^{\,2}+1\right)\,\left(\phi^{\,2}-\pi^{\,2}\right)+b_2\,\chi^{\,2}\right]^{\,2} \\ \nonumber
&+&
\frac{\mu^{\,2}}{2}\,\chi^{\,2}\,\left[(1-b_2)\,\left(\pi^{\,2}-\frac{\chi^{\,2}}{\gamma^{\,2}}\right)+b_2\,\phi^{\,2}\right]^{\,2}\,.
\een

In order to find analytical models consistent with the longitudinal movement of the molecular chain (which here is normalized to $\chi_v=\pm1$), we are going to constraint $\lambda$ and $\mu$ as
\be
\mu=\frac{\pi^{\,2}}{1+\pi^{\,2}}\,\lambda\,; \qquad \gamma= \frac{1}{\pi}\,.
\ee
Consequently, choosing $\lambda=7.30\,\pi^{\,-2}$ we find that $\mu\approx 0.67$, yielding to a family of physical potentials which can describe the twiston behavior naturally. The features of such potentials are unveiled in the graphics of Figs. \ref{FIG2} and \ref{FIG3}. There, we show how the $b_2$ parameter changes the shape of potential $V$. In Fig. \ref{FIG2} we built a graphic for $b_2=0$, which unveils the presence of nine vacua, whose four are connected by analytical orbits. The solid red orbit consist in the set of analytical solutions from Eq. $(\ref{twist02})$, while the dotted red orbit is simply given by
\be \label{twist11}
\phi(x)=-\pi \, \sqrt{\frac{1}{2} \left(1-\tanh \left(\pi ^2\, \mu\,  x\right)\right)}\,;\qquad \chi(x)=-\pi \, \gamma \sqrt{\frac{1}{2}\, \left(\tanh \left(\pi ^2\, \mu \, x\right)+1\right)}\,.
\ee
Due the symmetry of such a potential, the previous orbit also satisfies its equations of motion.  Besides, in Fig. \ref{FIG3} we plotted $V$ for $b_2=0.5$, unveiling a new set of vacua of our potential. Again, four of these vacua are connected by orbits $(\ref{twist02})$ and $(\ref{twist11})$. Moreover, we realized that the case $b_2=1$ is equivalent to the potential analyzed by Bazeia {\it et al.} in \cite{bazeia99,bazeia00}. Another interesting fact is that this new models present
\be \label{twist12}
E_{BPS}=\frac{\lambda}{4}\,\pi^{\,4}=17.99\,,
\ee
as its BPS energy, which is the same one derived in \cite{bazeia99,bazeia00}.

\section{Final remarks}
\label{remarks}

In this letter we applied the extension method to find analytical twistons for the crystalline polyethylene. For the first time, an analytical twiston model was constructed by a non-trivial combination of two one field systems. The analytical solutions used in this methodology can satisfy an entire family of two scalar field models, whose potentials do not present a line of zeros. Therefore, these new models agree with a physically consistent  longitudinal shift for the molecular chain, which should be proportional to the distance between adjacent molecular groups. Such a shift costs finite energy whose value is exactly the same computed in \cite{bazeia99,bazeia00}. We believe that the procedure presented in this letter can be applied to other interesting scenarios where the a two scalar fields model describes a molecular dynamics like polyaniline \cite{mge91}, leucoemeraldine \cite{mgme}  or fullerene \cite{bazeia02}. There are other models describing crystalline polyethylene based on sine-Gordon potentials, as the one investigated in \cite{hammad}, which could be revisited with this method. Moreover, we also intend to apply this procedure in the context of biological fluids, following the lines of \cite{sh12}, and in a three-field description of twistons as pointed in \cite{bazeia00,bazeia99}.

\acknowledgements
%{\bf Acknowledgements.} 

The authors would like to thank CNPq, and CAPES (Brazilian agencies) for support.

\end{document}